# A Novel Bio-Inspired Hybrid Multi-Filter Wrapper Gene Selection Method with Ensemble Classifier for Microarray Data

Babak Nouri-Moghaddam[1], Mehdi Ghazanfari[2]✉, Mohammad Fathian[3]

[a] Department of Industrial Engineering, Iran University of Science and Technology, Tehran, 1684613114, Iran, babaknouriit85@gmail.com
[b,*] Department of Industrial Engineering, Iran University of Science and Technology, Tehran, 1684613114, Iran, mehdi@iust.ac.ir
[c] Department of Industrial Engineering, Iran University of Science and Technology, Tehran, 1684613114, Iran, fathian@iust.ac.ir

## Abstract

Microarray technology is known as one of the most important tools for collecting DNA expression data. This technology allows researchers to investigate and examine types of diseases and their origins. However, microarray data are often associated with challenges such as small sample size, a significant number of genes, imbalanced data, etc. that make classification models inefficient. Thus, a new hybrid solution based on multi-filter and adaptive chaotic multi-objective forest optimization algorithm (AC-MOFOA) is presented to solve the gene selection problem and construct the Ensemble Classifier. In the proposed solution, to reduce the dataset's dimensions, a multi-filter model uses a combination of five filter methods to remove redundant and irrelevant genes. Then, an AC-MOFOA based on the concepts of non-dominated sorting, crowding distance, chaos theory, and adaptive operators is presented. AC-MOFOA as a wrapper method aimed at reducing dataset dimensions, optimizing KELM, and increasing the accuracy of the classification, simultaneously. Next, in this method, an ensemble classifier model is presented using AC-MOFOA results to classify microarray data. The performance of the proposed algorithm was evaluated on nine public microarray datasets, and its results were compared in terms of the number of selected genes, classification efficiency, execution time, time complexity, and hypervolume indicator criterion with five hybrid multi-objective methods. According to the results, the proposed hybrid method could increase the accuracy of the KELM in most datasets by reducing the dataset's dimensions and achieve similar or superior performance compared to other multi-objective methods. Furthermore, the proposed Ensemble Classifier model could provide better classification accuracy and generalizability in microarray data compared to conventional ensemble methods.

## 1 Introduction

One of the most important advances in medical technology in the recent decade is the development of DNA Microarray technology, which makes it possible to solve the problem of gene expression profiling [1]. Cancer occurs mainly due to the changes in genes and their unwanted mutations. Identification of these genes and their effects on the development of cancer plays a major role in the diagnosis and treatment of such diseases. One of the important applications of microarray data is its use in identifying cancer patients [2]. Accordingly, in recent years, extensive studies have been conducted by relying on machine learning and data mining methods to provide classification methods in cancer diagnosis using microarray data. However, using machine learning methods to classify such data is very challenging due to reasons such as a small number of samples, a great number of genes, imbalanced data, data complexity, and data shift [2].

The presence of a large number of genes makes it very difficult to identify and select genes that are effective in the disease. Also, it makes the classification models more complex and increases the training time of these models. Moreover, a small number of samples and the imbalanced data make classification models after training to be less generalizable to predict new data labels. To overcome the challenges of microarray data, solutions such as gene selection to reduce dimensions and ensemble learning to increase the generalizability of the classification model have been considered by many researchers [1–3]. The solutions of reducing dimension in microarray data are divided into two general classes of feature selection and feature extraction. They have positive results such as simplifying classification models, reducing learning time, and usually increasing classification accuracy. For dimension reduction, feature extraction methods maps the original gene space to lower dimension space [4]. However, this approach reduces the interpretability of the original genes and the possibility of identifying the effect of the original genes on the final result. Unlike the previous methods, feature selection aims at identifying a subset of prominent genes among all genes and uses statistical methods or meta-heuristic search methods to achieve this objective. In feature selection methods, subsets of selected genes are composed of genes of the main dataset. Therefore they allow researchers to have better interpretation and analysis [5, 6]. Thus, feature/gene selection is one of the suitable solutions to reduce the dimensions of microarray data and selects genes that are effective in different types of cancers. However, the problem of selecting the subset of prominent genes in microarray data is known as an *NP-hard* problem [7].

Gene selection methods can generally be divided into five classes, namely filter, wrapper, embedded, ensemble, and hybrid [2]. In filtering methods, the inherent relationship between genes and output is measured using statistical techniques and information theory, and related genes are identified. These methods include Correlation Based Filter[8], ReliefF[9], Symmetrical Uncertainty(SU) [8], Information Gain(IG) [4], and so on. In addition, these methods have low computational overhead and high scalability, but they might reduce the classification accuracy because of not using a classification model in the gene selection process. On the other hand, Wrappers include a meta-heuristic search methods (e.g. GA, FOA, and PSO) and a classification model that tries to identify a quasi-optimal subset of genes by considering the classification performance criterion. Wrappers usually have high computational overhead, but they provide better results than the filter due to using classifiers in the search process. Hybrid methods typically use a combination of a filter and a wrapper to select genes. Hybrid methods try to exploit the positive points of both filter and wrapper methods simultaneously so that the filter method, as a preprocessing step, can reduce the dimensions of the gene space and then the wrapper method can select prominent genes among the remaining genes [1, 4, 10].

Given the advantages of hybrid methods, they have been the subject of intense research in recent years. Based on the search method, hybrid methods can be divided into single-objective and multi-objective groups. Many hybrid single-objective methods such as GA [11, 12, 21, 13–20], PSO [22, 23], FOA [24], DE [25], ACO، [26, 27], Gravitational Search Algorithm [28], Shuffled Frog Algorithm [29], Cuckoo search [30] have been proposed so far. Additionally, hybrid multi-objective methods have been considered by many researchers in recent years for the simultaneously optimizing the objectives of minimizing the number of genes and maximizing the efficiency of the classification model. Some of these methods are NSGA-II [31–34], MOPSO [35], MOGA [36], MOSHO [37, 38], MOSSO [39], MOACO [40], and so on.

Imbalanced data and the small number of samples are other problems of Microarray data that challenge the performance of classification models in training and coping with unseen data. One method to cope with such challenges is to use ensemble learning models. The aim of developing such a system is to offer a trade-off solution between test error and training error in an automated classification model. Ensemble models have been used effectively so far in a range of problems like feature selection, missing features, imbalanced data, incremental learning, concept drift learning, and other applications [41]. The main difference among these methods stems from three factors: 1) the way of selecting the training data, 2) the process of creating ensemble learner members, 3) and the law of combining the output of classifiers [41]. The ability of ensemble classification models in reducing training error and increasing the generalizability of the model has made them usable in the area of feature selection studies [42–50].

The problem of gene selection usually follows two conflicting objectives, which include reducing the number of genes and increasing the efficiency of classification. Accordingly, it can be solved as a multi-objective optimization problem (MOP). The final solution for MOPs is usually presented in the form of a set of non-dominated solutions, which is a trade-off between conflicting objectives [51]. Also, given the research literature, studies on solving the problem of gene selection using the hybrid multi-objective method are more limited compared to those of the hybrid single-objective method. Most of the proposed hybrid solutions have solved the problem of gene selection only by considering the criterion of classification efficiency as a single-objective [2, 3, 5].

## 1.1 Objectives

Given what was stated above, to solve the problem of gene selection and to construct an ensemble classification model in Microarray data, a hybrid solution based on multi-filter and novel adaptive chaotic multi-objective forest algorithm (AC-MOFOA) wrapper method is presented in this paper. In the proposed solution, multi-filter is employed as a pre-processing step to reduce the dimensions of the dataset. Due to the combination of several filter methods (i.e., IG, mRMR, RelifF, CFS, and Fisher-score), the multi-filter method has less bias in selecting distinct genes compared to single-filter methods. Also, the multi-objective forest algorithm as a wrapper minimizes the number of genes, maximizes the classification criteria, and optimizes the ELM kernel parameters, simultaneously. Accordingly, the final output of the wrapper will be a set of non-dominated solutions that has solutions with different gene numbers and classification accuracy, and provides the diversity needed to construct an ensemble classification model. The results of the proposed algorithm are compared with five hybrid multi-objective methods (i.e., MOSSO, MOCEPO, C-HMOSHSSA, NSPSO, and MOBBBO) on nine public microarray datasets, which have the number of different genes, classes, and samples. The main objectives of this study are as follows:
- Providing a new solution based on multi-filter and wrapper for gene selection in the microarray
- Introducing a new adaptive multi-objective forest algorithm based on chaos theory and non-dominated sorting
- Using multi-filter to pre-process data and reduce the number of data genes
- Selecting effective genes simultaneously with optimizing KELM classifier parameters
- Constructing ensemble classifier using the results of the final Pareto front

The rest of this article is organized as follows: In Section 3, we will explain the fundamental principles of the theory and previous methods of gene selection. In Section 4, the proposed idea for solving the gene selection problem is described. Section 5 presents the design of the experiments and their details. In Section 6, the results of the experiments are analyzed. Finally, in Section 7, conclusions and suggestions for future works are presented.

# 2 Background

## 2.1 Forest Optimization Algorithm (FOA)

The search process in wrapper methods is usually based on a metaheuristic method such as GA, and PSO. The FOA is one of the new metaheuristic algorithms that has been introduced in recent years. This algorithm tries to provide a solution for optimization problems by modeling trees' reproduction, growth, and competition in the forest.

FOA has features such as high speed, low number of function evaluations, effective global, and local search. In recent years, this algorithm has been successfully used to solve the problem of feature selection as a single-objective approach [52, 53]. The general steps of FOA are as follows:

1- *Initial forest population formation*: randomly initialize each tree with $age = 0$.

2- *Local seeding operator*: At this stage, some trees are selected as parents among ones that their age is zero by methods such as the roulette wheel. This operator produces some children with the age of zero around each parent by applying a single change to parent variables. The number of children generated by this operator for each parent is known as Local Seeding Change (*LSC*). Besides, the age of all trees will be increased by one in exception of the best tree and new generated children which is considered zero in FOA.

3- *Population control*: To control the population of trees in the forest, FOA uses two parameters of *age* and *Area-Limit*. If the number of trees exceeds the *Area-Limit*, the trees that are older than the maximum allowable age are first removed from the forest and added to the candidate population. Next, if the number of trees is still more than *Area-Limit*, the trees are sorted in descending order based on the value of the objective function, are selected among the trees in the size of *Area-Limit*, and then are transferred to the next generation. Other remained trees are added to the candidate population.

4- *Global Seeding Operator*: To increase exploration capability in FOA, the global seedling operator is applied to trees in the candidate population. This operator randomly selects some trees among the trees of the candidate population and randomly initializes several variables in each tree. The number of changed variables and the number of selected trees in the global seedling operator are some FOA parameters that are denoted by Global Seeding Change (*GSC*) and Transfer rate, respectively.

5- *Termination condition*: Similar to other metaheuristic algorithms, one of the following conditions can be considered for termination of execution: 1) reaching the defined accuracy threshold, 2) reaching the specific iterations, or 3) reaching a specified number of function evaluation (*NFE*).

## 2.2 Extreme Learning Machine (ELM)

The ELM model, as one of the newest classification models based on artificial neural networks, was presented by Huang. ELM is based on Single-hidden Layer Feedforward Networks (SLFNs) [54]. This neural network model has been presented based on not adjusting the parameters of the hidden-layer network. Compared to conventional neural network training methods, ELM has features such as better generalizability, faster learning, and not being trapped in local optimum [55]. For linear separation of data and improving classification efficiency, Kernel ELM, as a version of ELM, maps data from a smaller space to a larger space. Based on the details provided in the Supplementary Information, the performance of KELM depends on the values of $\gamma$ of the kernel function and $C$ (i.e., the regularization parameter). Hence, these two parameters need to be optimized.

## 2.3 Filter Methods

Using statistical techniques and information theory, filter methods measure the intrinsic relationship between the genes and output and identify prominent genes. These methods have low computational overhead and high scalability. However, due to the lack of using a classification model in the gene selection process, they might compromise the accuracy of classification. Filter methods can be divided into two classes, including Univariate and Multivariate. Univariate methods independently examine the dependence of each gene on the target output, in which the relations between the genes are ignored. In contrast, multivariate methods try to reduce the redundancy of gene subsets by considering the dependence of each gene on the target output and relationships between the genes. Multivariate methods have a higher computational overhead compared to univariate methods. The description of the five filter method used in this paper (i.e., IG, mRMR, ReliefF, CFS, and Fisher-score) is presented in Supplementary Information.

## 2.4 Multi-Objective Optimization

Problems in the real world usually include the objectives that need to be optimized simultaneously. To solve such problems, a set of solutions that represents a tradeoff among different objectives is required. The set of solutions to multi-objective problems is known as Pareto optimal solutions. A multi-objective optimization (MOO) problem is defined as follows [56, 57]:

$$\min \vec{G} := \left[ G_1(\vec{v}), G_2(\vec{v}), ..., G_k(\vec{v}) \right] \quad (1)$$

*subject to:*

$$f_i(\vec{x}) \leq 0 \quad i = 1, 2, 3, ..., m \quad (2)$$

$$h_i(\vec{x}) = 0 \quad i = 1, 2, 3, ..., p \quad (3)$$

Where $\vec{v} = [v_1, v_2, ..., v_n]^T$ is a vector of decision variables, $G_i : R^n \rightarrow R$, $i = 1, ..., k$ is objective functions, and $f_i, h_j : R^n \rightarrow R$, $i = 1, ..., m$, $j = 1, ..., P$ represents constraint functions. The definitions of these parameters are as follows:

*Definition 1*: The vector $\vec{p} = (p_1, p_2, ..., p_k)$ dominates the vector $\vec{q} = (q_1, q_2, ..., q_k)$ as Pareto ($\vec{p} \preceq \vec{q}$), if the following relationship is established:

$$\forall i \in \{1, ..., k\} : p_i \leq q_i \; \wedge \exists i \in \{1, ..., k\} : p_i < q_i \quad (4)$$

*Definition 2*: The solution $\vec{v} \in \mathcal{F}$ ($\mathcal{F}$ represents the acceptable solution space) is an optimal Pareto solution if there is no solution $\vec{v}' \in \mathcal{F}$ in which $\vec{q} = \vec{G}(\vec{v}') = (\vec{G}_1(\vec{v}'), ..., \vec{G}_k(\vec{v}'))$ dominates $\vec{p} = \vec{G}(\vec{v}) = (\vec{G}_1(\vec{v}), ..., \vec{G}_k(\vec{v}))$.

*Definition 3*: For a given MOP, $\vec{F}(\vec{v})$, the optimal Pareto set $\mathcal{P}^*$ is defined as follows:

$$\mathcal{P}^* := \{\vec{v} \in \mathcal{F} \mid \neg \exists \vec{v}' \in \mathcal{F} \; \vec{G}(\vec{v}') \preceq \vec{G}(\vec{v})\} \quad (5)$$

*Definition 4*: For a given MOP, $\vec{G}(\vec{v})$, the optimal Pareto set $\mathcal{P}^*$, and optimal Pareto set, $\mathcal{PF}^*$ is defined as follows:

$$\mathcal{PF}^* := \{\vec{u} = \vec{G}(\vec{v}) \mid \vec{v} \in \mathcal{P}^*\} \quad (6)$$

As a result, the optimal Pareto front of the set $\mathcal{F}$ of all decision variable vectors will include members that meet the criteria 2 and 3.

## 2.5 Literature review

Gene selection is one of the important methods in the preprocessing of microarray data to reduce data dimensions and simplify classification models. In general, gene selection methods can be divided into 5 groups: Filter, Wrapper, Hybrid, Ensemble, and Embedded. Among the mentioned methods, hybrid methods have been considered by many researchers in recent years as they combine positive features of Filter and Wrapper methods. Filter methods use statistical models and information theory to identify genes related to the output. These methods have low computational overhead and high scalability. However, since they do not use a classification model, they usually cause a reduction in classification efficiency criterion. Wrapper methods use a search method to find subsets of quasi-optimal genes and use the accuracy/error of a classification model to evaluate the found subsets. Due to using a classification model in the search process, wrapper methods usually provide better results but they have a high computational overhead. Thus, in Hybrid methods, by combining the filter and wrapper methods, an attempt is made to use the advantages of both methods simultaneously in the final gene selection model. In hybrid methods, a filter is usually applied to the data as data preprocessing step to select the genes that have the most relationship with the output and the least relationship within the set, and to reduce the initial data set dimensions. Next, a wrapper is used to select a subset of quasi-optimal genes among the remaining gene sets. We review the hybrid methods of gene selection in this section. Depending on the type of search method, hybrid methods can be divided into single-objective and multi-objective groups. Hybrid single-objective algorithms usually follow the objectives of minimizing the number of selected genes or maximizing the efficiency of classification or a combination of these objectives.

Shreem et al. [58] proposed a hybrid solution based on Harmony search and the Markov blanket filter in which the Markov blanket was considered as a local search to improve harmony search solutions. In [27], a hybrid solution is presented based on the Fisher-score filter to reduce the initial dimensions and the ACO algorithm for selecting the optimal gene subset. Elyasigomari et al. (2017) proposed a hybrid solution based on the mRMR filter and the hybrid algorithm of the cuckoo optimization algorithm (COA) and harmony search [30]. In this method, the Harmony search method is used as a local optimizer to improve COA solutions. In [20], a hybrid method was proposed based on the T-test filter and the Nested-GA algorithm in which GA has two sections of outer and inner. The outer section selects genes based on the accuracy of the SVM classifier and the inner section is executed on the DNA Methylation dataset. Shukla et al. [19] proposed a new idea based on conditional MI and adaptive GA filters. In another study, a hybrid solution based on ensemble filter and GA was presented [18]. In this method, three filter methods of ReliefF, Chi-Square, and Symmetrical Uncertainty were used to construct the ensemble filter. Based on the conducted study, the Union merge function outperforms the Top N Gene. In [22], a method was proposed based on correlation base filter and improved PSO. Gangavarapu et al. [59] proposed a new idea based on an ensemble filter that uses 5 filter methods of mRMR, IG, CFS, CFSS, and oneRFeatureEval to form the ensemble filter. Then, the parameters of the penalty functions are optimized using Greedy search and GA. In [60], a hybrid solution is presented based on Artificial Bee Colony (ABC) called PrABC, which is an objective function as a combination of classification accuracy and the number of features. In this method, the dimensions of the dataset are reduced using an Ensemble Filter based on IG, Correlation, and Relief. Finally, PrABC is used to select the optimal gene subset. Bir-Jmel et al. [26] presented an innovative hybrid method using a filter based on the MWIS graph and ACO. In the proposed method, a local search is used to improve ACO solutions. In [24], a hybrid solution is proposed based on ANOVA Filter and enhanced Jaya-based FOA. In this method, both parameters LSC and GSC in the FOA are considered in the range of 1%-50% and optimized using enhanced Jaya. Other hybrid single-objective methods include GA [12–21], PSO [22, 23], FOA [24], DE [25], ACO [26, 27], Shuffled Frog Algorithm [29], Cuckoo search [30], and so on.

Multi-objective metaheuristic algorithms optimize several conflicting objectives in their search process [61]. Moreover, the problem of gene selection is inherently a multi-objective optimization problem that has at least two conflicting objectives: 1) minimizing the number of genes and 2) maximizing the criterion of classification efficiency. Many researchers have recently focused on solving the problem of multi-objective gene selection. The solution of such algorithms is a set of non-dominated solutions that provide a tradeoff among different objectives for users. The set of non-dominated solutions is also known as the Pareto Front solutions.

In [36], a hybrid solution is proposed based on the relevance filter and MOGA. In this method, first, the relevance filter selects 100 genes and then MOGA continues the search by considering the objectives of the accuracy of classification and size of the subset of selected genes. Li et al. [62] proposed a new idea by combining the Fisher Score-Markov filter and the Multi-Objective Binary Biogeography Based Optimization (MOBBBO) algorithm. In this approach, Fisher-Markov as a preprocessing step reduces dimensions of the database by removing irrelevant genes. Next, MOBBBO simultaneously uses the search to find the optimal subset of genes and optimize the parameters of the SVM kernel function. In another proposed solution, a combination of correlation coefficient and NSGA-II is presented [32]. In [35], a hybrid method is presented by considering Quartile filters and non-dominated sorting MOPSO.. Lai et al. proposed a hybrid idea based on the averaged filter method (AFM) and MOSSO [39]. AFM is an Ensemble Filter created by combining several filter methods. In [63], a model is proposed by combining filter and wrapper methods in which the T-test filter is considered as one of the objectives. Baliarsingh et al. [64] applied the Fisher-score method along with the Multi-Objective Chaotic Emperor Penguin Optimization (MOCEPO). In the proposed method, conventional methods of generating random numbers were replaced with chaos theory to improve efficiency. In [65], a solution called C-HMOSHSSA is presented by combining Fisher-score and multi-objective spotted hyena optimizer (MOSHO) and the Salp swarm algorithm (SSA). In another study, a combination of a Multi-filter and BOFS was used to solve the gene selection problem and construct an ensemble classifier [65]. In [66], a hybrid solution was presented based on Fisher-score and MOFOA in which the concepts of repository and binary tournament selection are used to solve the problem of multi-objective gene selection. Divya et al [37] have proposed a hybrid solution based on IG filter and MOSHO considering the accuracy of SVM and the number of the selected genes. Some other hybrid multi-objective methods include NSGA-II [31–34], MOPSO [35, 39, 67], MOGA [36], MOSHO [37, 38], MOSSO [39], MOACO [40], MOBAT [68, 69], and so on.

Most of the studies conducted to solve the gene selection problem have focused on single-objective metaheuristic algorithms. In this regard, studies conducted on hybrid multi-objective gene selection methods are less than single-objective studies. Thus, our study aims to present a hybrid multi-filter and multi-objective wrapper solution based on the Forest Optimization Algorithm to solve the gene selection problem and construct an ensemble classifier.

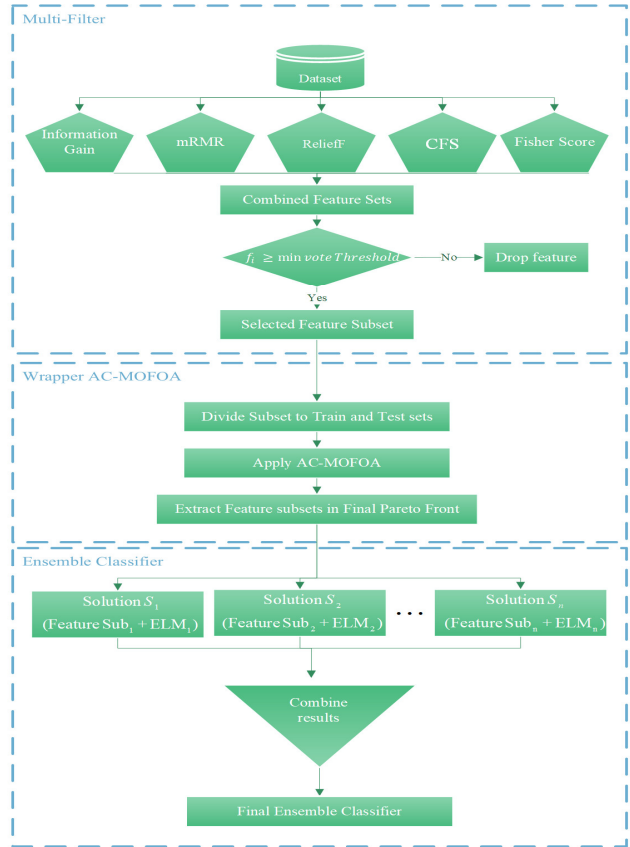

**Fig. 1** The flowchart of the proposed method

## 3 The proposed idea

Given what was stated in the previous section, it can be concluded that the problem of gene selection is a multi-objective problem that must simultaneously solve two conflicting objectives of reducing the number of genes and increasing the efficiency of the classification model. The FOA algorithm is one of the latest metaheuristic algorithms used as a single-objective and multi-objective to solve the problem of feature/gene selection [24, 53, 66]. However, based on the "no free lunch" theory, no solution is optimal for all problems and a new and improved solution can always be proposed. Hence, this study presents a new solution based on multi-filter and AC-MOFOA to simultaneously solve the problem of gene selection and the construction of ensemble classifiers.

In the proposed solution, the Multi-filter first reduces the dimensions of the dataset by removing Irrelevant and redundant genes. The considered multi-filter consists of 5 filter methods of IG, Fisher-score, mMRM, CFS, and ReliefF. AC-MOFOA is then presented as a new version of the FOA algorithm based on the concepts of chaos theory, adaptive operators, non-dominated sorting ($NS$), and crowding distance. AC-MOFOA identifies quasi-optimal gene subsets, considering the objectives of maximizing KELM classification accuracy and minimizing the number of selected genes. Based on the concepts of MOO, the output of MOP problem-solving algorithms is presented as a set of solutions that are not superior to each other. Also, the members of the solutions set have the necessary variety to construct Ensemble classifiers due to training with different subsets of datasets and different KELM classifiers. Therefore, members of the final Pareto Front set are used to construct the Ensemble classifier to obtain a final classification model for microarray data classification. To have a good understanding, the general flowchart of the proposed method is shown in Figure 1, followed by describing the main sections.

### 3.1 Multi-Filter step

Filter methods have low computational load and high scalability. However, each filter approach has advantages and disadvantages and using only one filter method to identify related genes can cause bias in the final result. Thus, in the proposed method, a multi-filter is used for the initial preprocessing of the data and identification of subset of genes that have the highest correlation with the output and lowest internal correlation. The objective of multi-filter is to reduce the bias effect of filter methods in gene selection and to combine the strengths of different filter methods. The desired multi-filter includes five filters, including IG, Fisher-score, mMRM, CFS, and ReliefF. In this system, Fisher-score and IG are univariate filter methods and mMRM, CFS, and ReliefF are multivariate filter methods. In this step, each of the five methods in the multi-filter is first applied to the dataset and then 30% of the superior genes are extracted from the output of each method based on the rank of the genes. Then, using the voting mechanism and "min vote" threshold, the results of five filter methods are combined and the subset of output genes of the multi-filter step is determined. In this study, the min vote threshold was considered to be 3.

## 3.2 Wrapper AC-MOFOA step

In the present study, a new version of the FOA called AC-MOFOA is proposed. This new algorithm is based on the concepts of chaos theory, adaptive operators, non-dominated sorting, and crowding distance. One of the important challenges in FOA is determining the optimal value of *LSC* and *GSC* parameters. Thus, adaptive local seeding and adaptive global seeding operators were proposed to automatically adapt the *LSC* and *GSC* values during the search process to solve this challenge in AC-MOFOA. Chaos theory was also used to improve AC-MOFOA diversity and faster convergence. AC-MOFOA optimizes KELM model parameters in addition to searching for quasi-optimal gene subsets. AC-MOFOA pseudo code is displayed in Algorithm 1, with its steps described in the following.

---

**Algorithm 1: AC-MOFOA pseudo code**

1: **begin**
2:  Apply Multi-Filter
3:  get most discriminative genes from Multi-Filter
4:  Divide each Microarray Dataset into two sets (i.e., Training set and a Test set)
5:  Initialize the Forest with Chaotic Trees
6:  Assess each Tree in the Forest according to objective functions
     (i.e., classification accuracy and selected features ratio)
7: **for** *T* =1 to *Max Number of Iterations* **do**
8:   Apply Non-Dominated Sorting to Forest
9:   *F1* ← Extract Lowest Rank Pareto Front according to Non-Dominated Sort results
10:  Set *Age* of all trees in *F1* Zero
11:  Calculate crowding distance for trees in *F1* according to equation (9)
12:  Calculate *LSC* according to equation (10)
13:  **for j = 1 to *Area-Limit***
14:    Apply *Roulette wheel* to select Parent Tree from *F1*
       /* the solution with more crowding distance have higher selection probability */
15:    Apply ***Adaptive Chaotic Local-Seeding***(*selected Parent, T, LSC*)
16:    Add new Trees to the Forest
17:  **end for**
18:  For All Trees in the Forest *Age* ← *Age* + 1  /*except for new trees add by local seeding operator*/
19:  Remove All trees with Age > *Max-Age* and them to ***Candidate population***
20:  **if** Forest size > Area-limit
21:      Remove trees from Forest according to Pareto rank and Crowding-distance
           and them to ***Candidate population***
       /* trees in Highest Pareto rank with lowest crowding distance will have high priority to remove */
22:  **end if**
23:  Randomly select some trees of the ***Candidate population*** according to the "***transfer rate***"
12:  Calculate *GSC* according to equation (11)
23:  **for each selected tree from *Candidate population***
24:    Apply ***Adaptive Chaotic Global-Seeding***(*selected tree, T, GSC*)
25:    Add new Trees to the Forest
26:  **end for**
27:  *T* ← *T* + 1
28: **end for**
29: For the solutions in *F1*, calculate the classification accuracy on the *Test* set
30: Return the solutions in *F1* as final *Pareto front* with their *Training* and *Test* classification accuracy
31: **End**

### 3.2.1 Initialization

In AC-MOFOA, each tree is denoted as a vector that represents the values of tree age, $C$ and $\gamma$ parameters related to KELM, and values related to gene selection $(i.e., v_1, v_2, \cdots, v_n)$. In this method, the size of each tree is $n + 3$, which includes $n$ number of data genes, and three variables including age, $C$, and $\gamma$. Figure 2 illustrates the structure of a tree or solution.

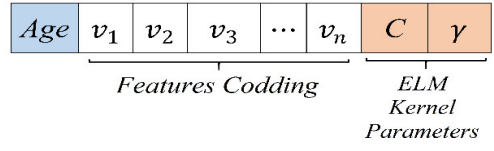

**Fig. 2** Tree structure in AC-MOFOA

AC-MOFOA uses Chaos theory to initialize the trees. Chaos theory is one of the new techniques used to improve the search ability of metaheuristic algorithms. Chaos can be described as the behavior of a non-linear dynamic system which is highly sensitive to the initial state and can be calculated by deterministic algorithms. In addition to having random features, chaos theory has other features such as sensitivity, stochasticity, and ergodicity to preliminary situations. Based on these features, chaos theory ensures diversity among the population-based metaheuristic algorithm solutions and enhances the convergence performance of these algorithms[64, 70, 71]. For this purpose, the logistic map function will replace the generation of random numbers. The logistic map functions have main behaviors: chaotic and convergent. The chaotic and convergent behavior leads to exploration and exploitation, respectively [72]. The logistic map function is defined as follows:

$$v_i^{t+1} = 4 \bullet v_i^t \bullet (1 - v_i^t) \qquad (7)$$

where $v_i^t$ represents the value of chaotic map for $i$-th variable, in which $t$ and $t+1$ represent he iteration number. Note that $v_i^0$ is randomly initiated in the interval $(0,1)$. In the next iteration of AC-MOFOA to calculate $v_i^2$ using seeding operators, the equation (7) will applied on $v_i^1$. In the proposed solution, based on the KELM studies, the parameters $C$ and $\gamma$ are considered in range of $[2^{-8}, 2^8]$ [73]. Due to initializing the values of $C$ and $\gamma$, $v_i^t$ in the equation (7) is replaced with $C^t$ and $\gamma^t$ where the results of this equation mapped to $[2^{-8}, 2^8]$. The value of the age variable will be zero for all new trees generated by AC-MOFOA operators.

The values of the variables $v_1, v_2, \cdots, v_n$ belong to an interval of $(0,1)$, which are mapped to binary space based on the equation (8). In equation (8), if the value of the $v_i$ is greater than 0.5, the binary value of the gene will be equal to 1; otherwise, it will be equal to 1. A value of 1 means that the gene is selected while a value of 0 means that it is not selected.

$$f(v_i) = \begin{cases} 1 & if\ v_i \geq 0.5 \\ 0 & otherwise \end{cases} \qquad (8)$$

### 3.2.2 Handling Multi-Objective

To solve the problem of multi-objective gene selection, some changes must be applied to FOA. For this purpose, the idea of *NS* and crowding distance presented in NSGA-II [74] was employed. *NS* is a technique for ranking Pareto optimal solutions based on the concept of dominance. *M* and *n* are the number of objectives and solutions, respectively. *NS* first extracts the set of solutions that are not dominated by any of the solutions obtained and forms the Pareto front $F_1$. Then, among the remaining solutions, the solutions that were dominated only by members of $F_1$ are extracted to form $F_2$ and this process continues accordingly. The set of solutions $F_1$ is considered as the Pareto optimal solutions. Crowding-distance is used in NSGA-II to prioritize members within each $F_i$ and direct search to less crowded areas. Crowding-distance is calculated using equation (9):

$$cd_i^m = cd_i^m + \frac{obj_m^{i+1} - obj_m^{i-1}}{obj_m^{max} - obj_m^{min}} \qquad (9)$$

where $m = 1, 2, \cdots, M$ is the number of objectives, $i$ is the number of the solution in the list sorted by *m*-th objective, and $obj_m^i$ represents the value of the *m*-th objective function for the solution *i*. The crowding-distance value for the boundary points is assumed to be $\infty$. A smaller value of $cd_i$ means that there are other solutions adjacent to the solution *i*, and the solution *i* is in a more crowded space, and vice versa. To direct the search to less crowded areas, sorting within each $F_i$ will be done in a descending order from larger *cd* values to smaller values.

### 3.2.3 Adaptive Chaotic Local-seeding

A local seeding operator has been introduced in the FOA to generate exploitation capability. It generates an *LSC* number of new trees for each parent around trees with an age of *zero*. The local seeding operator in AC-MOFOA is a single-parent operator. Each parent is selected among the $F_1$ members using the roulette wheel. In this method, members with larger *cd* values are more likely to be selected to increase the diversity between solutions. This operator changes the value of a variable in the parent tree using logistic map function according to equation (7). An example of this operator is illustrated in Figure 3.

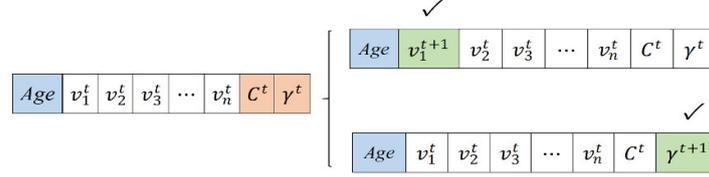

**Fig. 3** Example of a local seedling operator with $LSC = 2$

Determining the *LSC* value is one of the challenges of FOA and is determined by trial and error based on the dimensions of the problem. In the proposed solution to overcome this challenge, the *LSC* value is determined adaptively by the algorithm itself. Accordingly, *LSC* value is first set at 20% of the data set dimensions and then it is updated in each step through the following equation.

$$LSC = \left\lceil \frac{LSC}{e^{1/T}} \right\rceil \quad (10)$$

where *T* represents the current number of iterations of the algorithm. Based on equation (10), the minimum *LSC* value will be 1.

### 3.2.4 Population Limitation

Forest limiting is one of the most important operators of the FOA. In this process, old and improperly fitted trees are removed from the Forest and added to the candidate population. After implementing the local seeding operation, the age of all the trees in the forest increases by one unit. In AC-MOFOA, the age of $F_1$ member trees is considered to be 0. To limit the population, trees older than *Max-Age* are removed from the forest and added to the candidate population. If the forest size is still larger than the *Area-Limit* after removing the old trees, the trees with the highest Pareto rank will be removed from the forest and added to the candidate population. Additionally, if there is a need to remove a part of $F_i$, trees based on *cd* value (from lower value to higher value) are removed from $F_i$ and added to the candidate population.

### 3.2.5 Adaptive Chaotic Global-seeding

The global seedling operator is used to create exploration capability to generate new trees in FOA. This operator selects based on the size of the transfer rate parameter from the trees in the candidate population and then changes the *GSC* number of the variables in each tree. To change variables, a chaotic number in the allowable range of the variable is generated using equation (7) and replaces its previous value. Figure 4 illustrates the function of this operator.

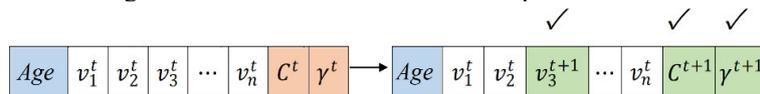

**Fig. 4** Global seedling operator performance in AC-MOFOA with $GSC = 3$

Another important parameter of FOA is the *GSC* parameter. Setting its optimal value is one of the challenges of FOA. In the proposed solution, the *GSC* value is set adaptively by the algorithm itself to overcome this challenge. In this method, the *GSC* value is first set at 30% of the dataset dimensions and then is updated in each step through equation (11):

$$GSC = \left\lceil \frac{GSC}{e^{1/T}} \right\rceil + 1 \quad (11)$$

Based on equation (11), the minimum value of *GSC* will be 2.

### 3.2.6 Objective functions

AC-MOFOA has two objectives of maximizing the accuracy of the classification and minimizing the number of genes. The classification accuracy is calculated using equation (12):

$$Acc = \frac{TP + TN}{FP + FN + TP + TN} \quad (12)$$

The parameters *FP*, *TP*, *FN*, and *TN* represent false positive, true positive, false negative, and true negative, respectively.

### 3.3 Creation of Ensemble classifier

Imbalanced data and the small number of samples are other issues of microarray data that challenge the performance of classification models in training and coping with unseen data. One way to cope with such challenges is to use ensemble learning models. The aim of developing such a system is to provide a trade-off solution between bias error (training error) and variance error (test error) in an automated classification system. One of the important features in constructing an ensemble model is to create diversity in basic classifiers. Several methods have been proposed to create diversity in basic classifiers; e.g. the use of different classification models or training each classification model with different subsets of datasets. The final solution of AC-MOFOA is a set of non-dominated solutions, each including a different subset of genes and a KELM with different kernel parameters. Therefore, it can be stated that the members of the final Pareto front have sufficient diversity in terms of basic classifiers and training subsets to construct an ensemble classifier. Accordingly, the output of the $F_1$ members is combined using the voting mechanism to form the final ensemble classifier.

## 4 Experimental Design

To design the experimental scenarios, nine microarray datasets [75, 76] were used considering the diversity in the number of samples, genes, and classes (Table 1). After reducing the dimensions in the multi-filter step, each of the datasets is randomly divided into two batches (80% for training and 20% for the test) while maintaining the ratio of classes. To evaluate the subset of selected genes, the KELM classifier with RBF kernel function is used and the kernel $(\gamma)$ and ELM $(C)$ parameters are optimized by the proposed solution. KELM and 10-fold cross-validation (CV) were used to evaluate the subset of selected genes [77].

Table 1 Selected datasets summary

| Datasets | # of genes | # of Classes | # of Samples |
|---|---|---|---|
| SRBCT | 2308 | 4 | 83 |
| Tumors_9 | 5726 | 9 | 60 |
| Leukaemia3 | 7129 | 3 | 72 |
| Colon_Prostate | 10937 | 2 | 355 |
| Lung | 12601 | 5 | 203 |
| GCM | 16064 | 14 | 190 |
| Breast | 24482 | 2 | 97 |
| Rsctc_5 | 54614 | 4 | 89 |
| Rsctc_6 | 59005 | 5 | 92 |

In this study, five Hybrid multi-objective methods, namely C-HMOSHSSA [38], MOBBBO [62], MOCEPO [64], MOSSO [39], and NSPSO [35], were selected for comparing and evaluating the AC-MOFOA results. Python 3.7 programming language is used to implement all methods. We also used the *ELMClassifier* function in the *Python-ELM v0.3* Library with default settings. A PC with 16GB ram and Intel Core i7 6700HQ hardware was used for performing the tests. To compare the results fairly, the parameters of the selected algorithms are obtained based on the information of reference articles or through the Taguchi experiment design method [78]. In all algorithms, the number of individuals (i.e. tree, habitat, hyena, salp, and particle) is equal to 60 and the termination condition is considered to be 40000 evaluations. The results were also examined in 50 independent performances. In algorithms with continuous expression, the threshold for selecting or not selecting a feature is considered to be 0.5. Table 2 presents a summary of the parameters of compared algorithms.

Table 2 Parameters and settings of selected algorithms

| Algorithms | Representation | Operators | Parameters |
|---|---|---|---|
| MOBBBO | Binary | Basic Habitat Migration and Mutation Strategy | *Habitat modification probability = 1, mutation rate = 0.5* |
| C-HMOSHSSA | Continuous | Standard Spotted Hyena and salp swarm operators | $c_1, c_2, c_3$ are random numbers, L is max iteration allowed, $g > 1$ |
| MOCEPO | Continuous | Standard Emperor Penguin operators | $M = 2$, $r_1, r_2$, are logistic map numbers |
| MOSSO | Continuous | Standard SSO operators | $c_g < c_p < c_w$, are random numbers |
| NSPSO | Continuous | Standard PSO operators | $r_1, r_2$, are random numbers, $0 < \alpha < 1$ |
| AC-MOFOA | Continuous | Adaptive Chaotic Local and Global seeding | *transfer-rate*=30%, *lifetime*=15, *LSC*=20%, *GSC* = 30% of variables |

A total of 50 executions were used for comparing the results based on the Pareto front. Accordingly, the obtained Pareto sets in 50 independent executions are first placed in a Union set and then non-dominated solutions are extracted among them to form the final Pareto set [61]. Also, in the next section, Box-plot charts are used to represent some of the obtained results. These charts were plotted based on the data collected in the Union set.

One of the quantified criteria for comparing the performance of multi-objective algorithms is the success counting criterion (SCC) [79]. Based on this criterion, to quantify the performance of multi-objective algorithms, the final Pareto front of each method is aggregated in set $P$ and then the Pareto front obtained by all methods is extracted from the members of set $P$. After this step, the level of contribution of each method in the formation of the optimal Pareto front is calculated using equation (13).

$$SCC = \sum_{i=1}^{n} S_i \quad (13)$$

In equation (13), if the solution $i$ belongs to the studied method, $S_i = 1$; otherwise, $S_i = 0$. Also, $n$ represents the total number of members of the optimal Pareto front. According to this criterion, a method with higher SCC values shows better performance in identifying the final Pareto front. Diversity and convergence measures are used to evaluate the performance of multi-objective metaheuristic algorithms in identifying the optimal Pareto front. The diversity measure assesses the level of diversity of Pareto front solutions and the convergence measure assesses the degree of convergence of solutions to the main Pareto front. Hypervolume is one of the criteria used for evaluating the performance of multi-objective metaheuristic algorithms. This criterion simultaneously evaluates both diversity and convergence measures [80, 81]. Hypervolume is calculated based on the following equation:

$$HV = volume\left(\bigcup_{i=1}^{|P|} y_i\right) \quad (14)$$

## 5 Experiments and Results

In this section, the test results are presented in five subsections: 1) providing multi-filter step results in terms of classification accuracy and dimension reduction rate, 2) comparison of AC-MOFOA with multi-objective algorithms, 3) AC-MOFOA performance evaluation using Student T-test, 4) presenting the comparison results of the examined methods in terms of CPU execution time, and space and time complexity, and 5) analyzing the results of the ensemble classifier. Each of the subsections is described in the following.

Table 3 Result of Applying Multi-Filter Step

| Datasets | # of genes | KELM acc. | # of selected genes by Multi-filter | % gene reduction ratio | KELM acc. On selected genes |
|---|---|---|---|---|---|
| SRBCT | 2308 | 88.23 | 472 | 79.5% | 90.72 |
| Tumors_9 | 5726 | 78.98 | 1209 | 78.8% | 84.41 |
| Leukaemia3 | 7129 | 97.22 | 1836 | 74.2% | 97.66 |
| Colon_Prostate | 10937 | 95.74 | 2081 | 80.9% | 97.89 |
| Lung | 12601 | 90.24 | 2745 | 78.21% | 93.97 |
| GCM | 16064 | 70.53 | 3653 | 77.25% | 78.03 |
| Breast | 24482 | 68.04 | 5992 | 75.52% | 86.53 |
| Rsctc_5 | 54614 | 69.66 | 14507 | 73.43% | 72.38 |
| Rsctc_6 | 59005 | 81.523 | 15511 | 73.71% | 82.69 |

### 5.1 Analyzing the multi-filter results

The Multi-Filter step is the first step in the proposed hybrid solution for pre-processing and reducing the data dimensions. Thus, we first analyze the results of this step. Based on Table 3, multi-filter consists of five filter methods of IG, Fisher-score, mMRM, CFS, and ReliefF. It could reduce the number of genes in all datasets by at least 73%. The highest reduction in the number of genes was observed in the Colon-Prostate dataset with 80.9% and the lowest reduction was observed in Rsctc_5 dataset with 73.43%. Also, comparing KELM accuracy results in two modes of training with all genes and training with selected genes suggests an improvement in classification accuracy for selected genes (Table 3). Due to employing the combination of Univariate and Multivariate filter methods, Multi-Filter ability and efficiency increased in selecting the most prominent genes from the dataset; also the bias of using a single filter method can be decreased. Thus, it can be concluded that the Multi-Filter step as preprocess step could effectively reduce dimensions of the datasets, and improve the classification accuracy.

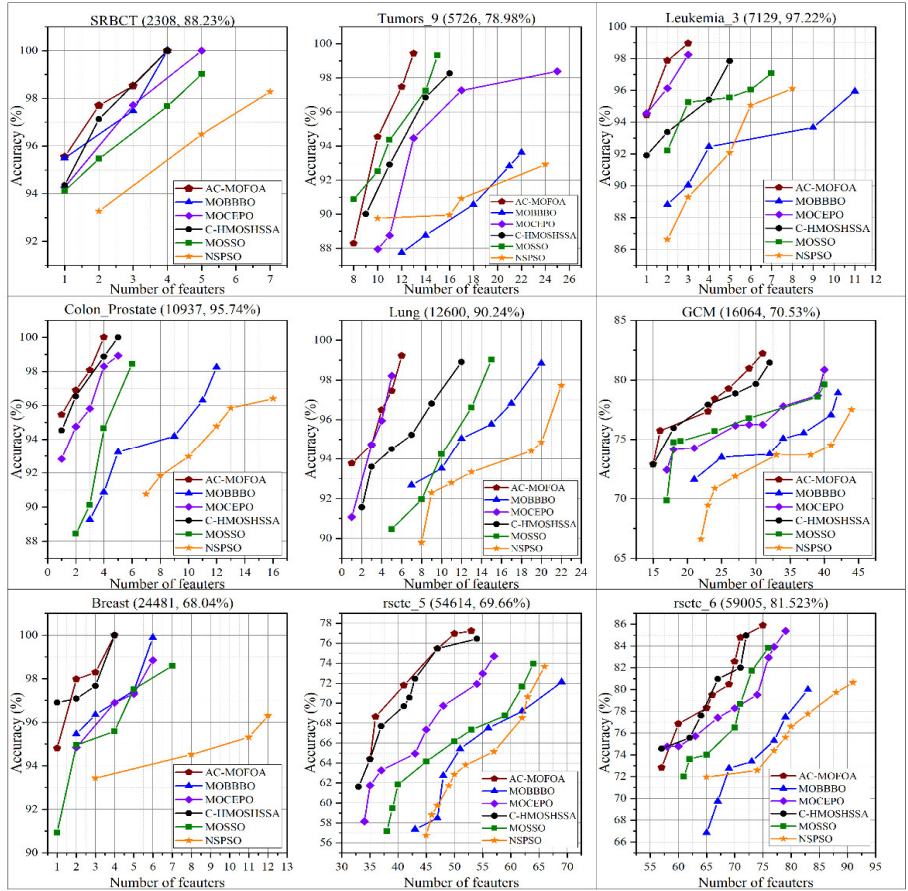

**Fig. 5** Comparing AC-MOFOA with other multi-objective algorithms based on non-dominated solutions on the test set

### 5.2 Analyzing of the AC-MOFOA results

To evaluate the performance of AC-MOFOA and compare its results with C-HMOSHSSA, MOBBBO, MOCEPO, MOSSO, and NSPSO methods, Figures 5-9 were plotted based on the final Pareto front, classification accuracy distribution, and distribution of selected gene numbers in two modes of test and train. Based on Figure 5, AC-MOFOA in 7 datasets could achieve higher classification accuracy by selecting fewer genes, and AC-MOFOA in 2 datasets (including SRBCT and Breast) showed performance similar to C-HMOSHSSA in terms of set objectives. Moreover, AC-MOFOA in all datasets could significantly increase the classification accuracy by considerably reducing the number of genes. For example, in the Breast dataset, AC-MOFOA by selecting 4 of 5592 genes (a set of genes selected by the Multi-filter) could achieve a classification accuracy of 100%.

To compare the level of contribution of each method in the formation of the final Pareto front, the SCC results on the test dataset were displayed in Table 4. Based on Table 4, in most datasets AC-MOFOA could achieve higher SCC values than other methods, suggesting that AC-MOFOA has a greater contribution in identifying the final Pareto front.

Table 4 SCC measure Comparison on the test set

| Datasets | AC-MOFOA | MOCEPO | MOBBBO | C-HMOSSA | MOSSO | NSPSO |
|---|---|---|---|---|---|---|
| SRBCT | 4 | 0 | 2 | 1 | 0 | 0 |
| Tumors_9 | 3 | 0 | 0 | 0 | 1 | 0 |
| Leukaemia3 | 3 | 1 | 0 | 0 | 0 | 0 |
| Colon_Prostate | 4 | 0 | 0 | 0 | 0 | 0 |
| Lung | 4 | 2 | 0 | 0 | 0 | 0 |
| GCM | 6 | 0 | 0 | 3 | 0 | 0 |
| Breast | 3 | 0 | 0 | 2 | 0 | 0 |
| Rsctc_5 | 6 | 0 | 0 | 4 | 0 | 0 |
| Rsctc_6 | 6 | 1 | 0 | 4 | 0 | 0 |

Metaheuristic algorithms have stochastic nature and achieve variant results in different executions. Thus, to compare the results of these algorithms, it is better to compare the distribution of solutions obtained in all 50 executions. Figure 6 presents the distribution of the accuracy of the solutions obtained by AC-MOFOA in the test data. Based on this figure, the accuracy of AC-MOFOA solutions in 50 independent executions in terms of statistical criteria such as mean and median in most datasets is better than that of other hybrid multi-objective methods. AC-MOFOA also outperformed other methods in terms of maximum and minimum accuracy. AC-MOFOA could increase the efficiency of KELM classification by optimizing the RBF kernel $(\gamma)$ and ELM $(C)$ parameters simultaneously with reducing the number of genes.

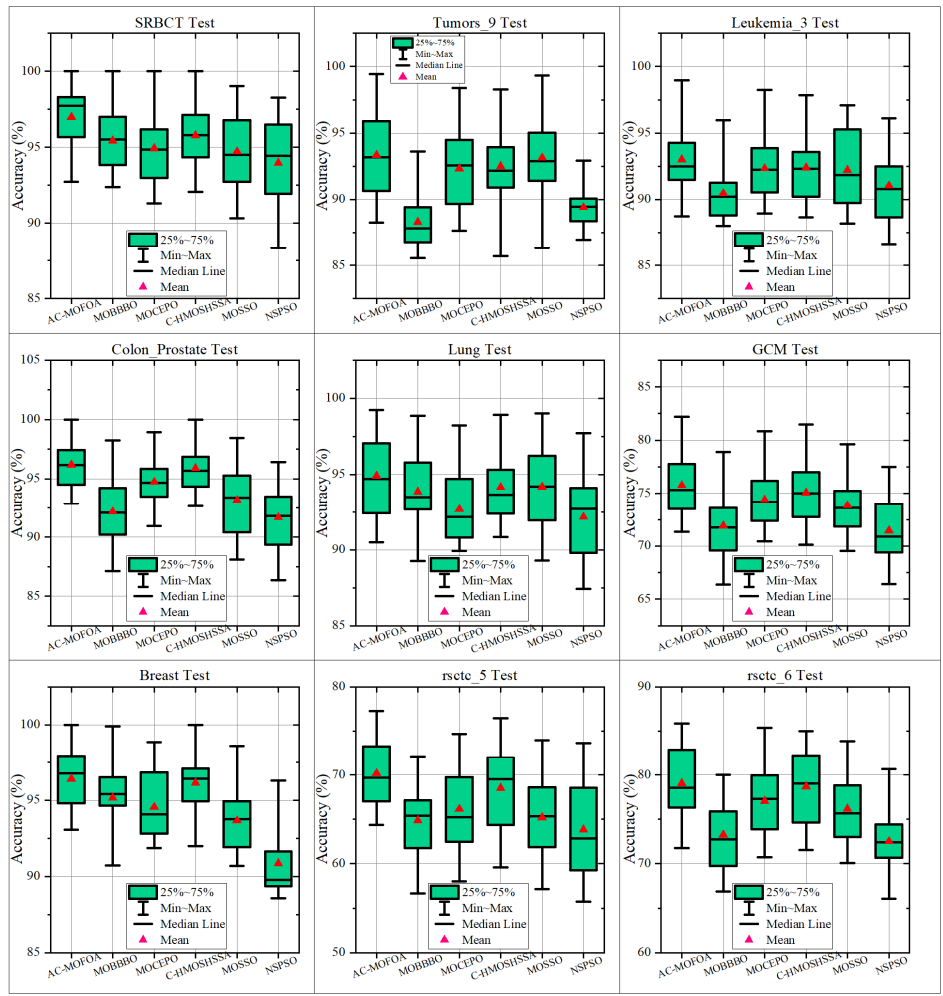

Fig. 6 Comparison of the distribution of classification accuracy of AC-MOFOA solutions on the test set

Figure 7 compares distributions of the selected genes numbers by AC-MOFOA and other multi-objective methods in 50 independent executions. Based on Figure 7, it can be seen that AC-MOFOA outperforms other multi-objective algorithms in most datasets in terms of the maximum and the minimum number of selected genes. Also, in 7 cases of datasets, the proposed method could provide better results than other methods in terms of the mean and median number of selected genes. Only in rsctc_5 and Tumors_9 datasets, C-SHMOSHSSA and MOSSO methods, respectively, outperformed AC-MOFOA in the objective function of minimizing the number of genes.

When comparing the distribution two objectives, they were examined independently and were different from the results in which both of these objectives were examined simultaneously using the concepts of dominance. For example, an algorithm may have achieved solutions with a higher number of genes but lower classification accuracy, while this solution has been dominated by other solutions in that algorithm.

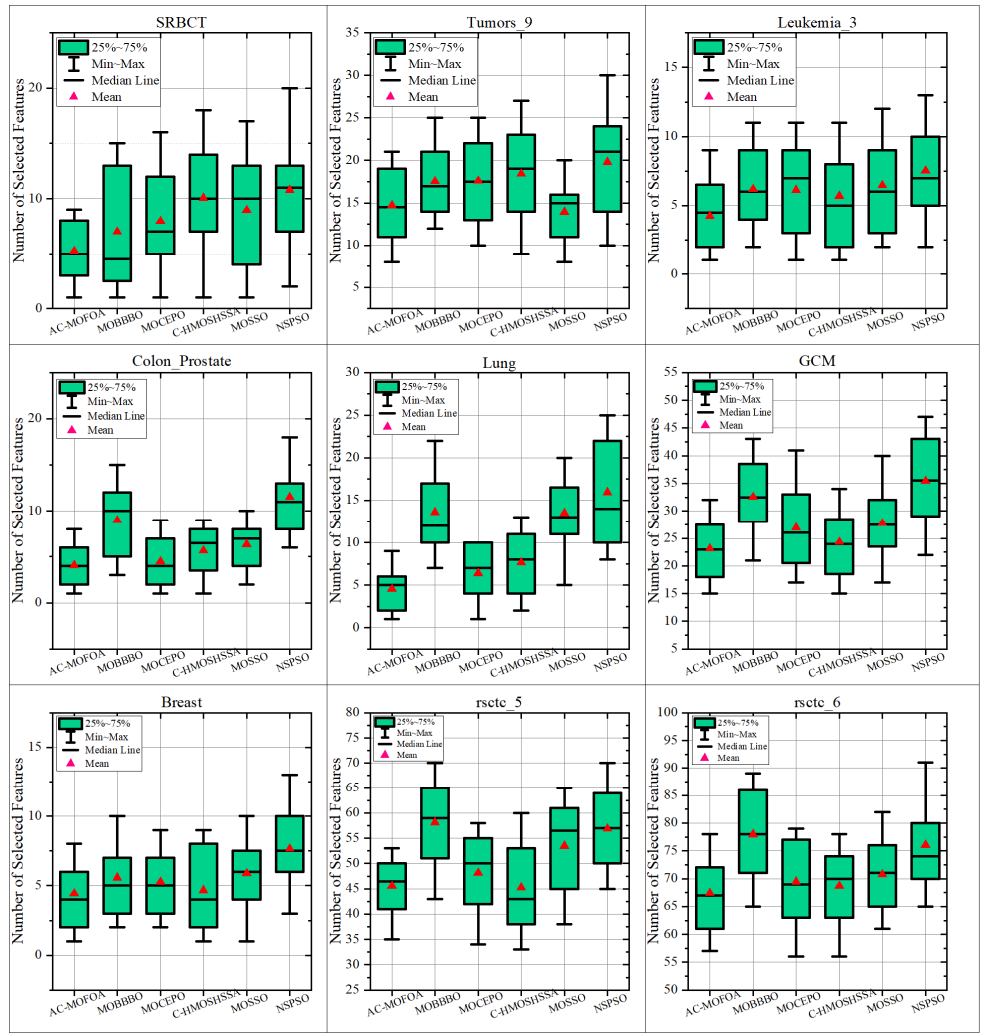

**Fig. 7** Comparing the distributions of the number of selected genes by AC-MOFOA

Figure 8 presents the results of comparing the AC-MOFOA Pareto front with the other 5 methods on the train set. According to this figure, AC-MOFOA solutions in most datasets could dominate the solutions of other methods. However, in the Breast dataset, C-HMOSHSSA achieves similar or better solutions compared to AC-MOFOA. Analysis of the SCC criterion in the training data is shown in Table 5. According to this table, AC-MOFOA had more contribution in identifying the final Pareto front and could identify more than 50% of the final Pareto front in 8 datasets.

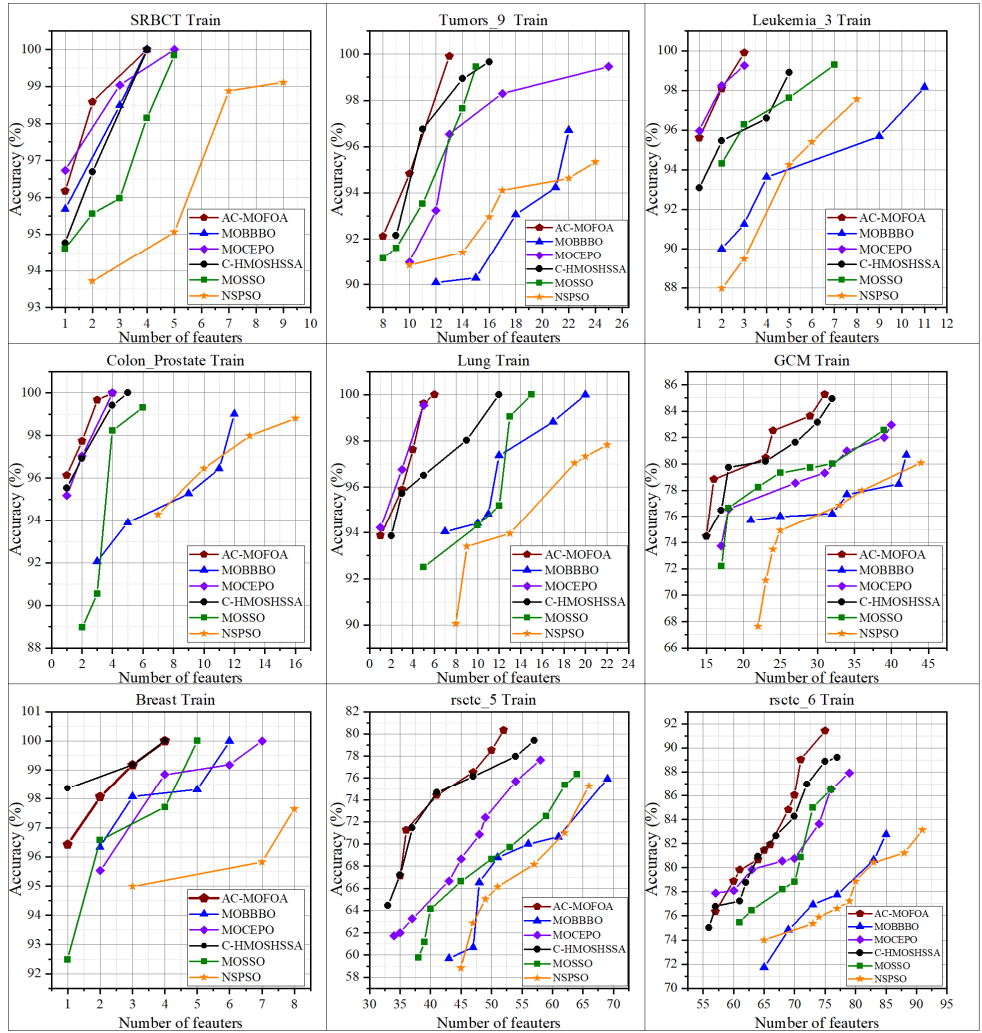

**Fig. 8** Comparison of AC-MOFOA with other multi-objective algorithms based on non-dominated solutions on the train set

Table 5 SCC measure Comparison on the train set

| Datasets | AC-MOFOA | MOCEPO | MOBBBO | C-HMOSSA | MOSSO | NSPSO |
|---|---|---|---|---|---|---|
| SRBCT | 2 | 1 | 0 | 0 | 0 | 0 |
| Tumors_9 | 3 | 0 | 0 | 1 | 0 | 0 |
| Leukaemia3 | 2 | 2 | 0 | 0 | 0 | 0 |
| Colon_Prostate | 4 | 1 | 0 | 0 | 0 | 0 |
| Lung | 3 | 2 | 0 | 0 | 0 | 0 |
| GCM | 6 | 0 | 0 | 2 | 0 | 0 |
| Breast | 2 | 0 | 0 | 3 | 0 | 0 |
| Rsctc_5 | 6 | 0 | 0 | 4 | 0 | 0 |
| Rsctc_6 | 9 | 1 | 0 | 3 | 0 | 0 |

Figure 9 shows the results of comparing multi-objective algorithms based on the classification accuracy distribution on the train set. As can be seen from this figure, AC-MOFOA in all cases could achieve higher or similar accuracy to other methods. Furthermore, AC-MOFOA always outperforms other methods in terms of minimum, mean, and median of accuracy.

The obtained results indicate that AC-MOFOA outperformed other methods in terms of achieving the optimal Pareto front and involving in the formation of the final Pareto front. In detail, high initial values of *LSC* and *GSC* in the adaptive local and global seeding operators lead to enhance the exploitation and exploration capabilities of the AC-MOFOA. Local seeding operator is applied to the solution of $F_1$ that boosts the ability of AC-MOFOA to improve the Pareto solutions further. Moreover, employing the global seeding operator on the Candidate Population gives a chance to less fitted trees for better exploration of the problem space. In addition, AC-MOFOA uses $NS$ to identify the Pareto front and rank the solutions. Furthermore, AC-MOFOA implements the crowding distance, which increases its ability to direct the search to the less crowded areas of the Pareto front, leading to a more uniform final Pareto front. Also, a separate study on distribution of each objective (namely, classification accuracy and the number of selected genes) shows that AC-MOFOA solutions were superior to other methods in most datasets in terms of statistical criteria such as upper bound and lower bound, mean and median.

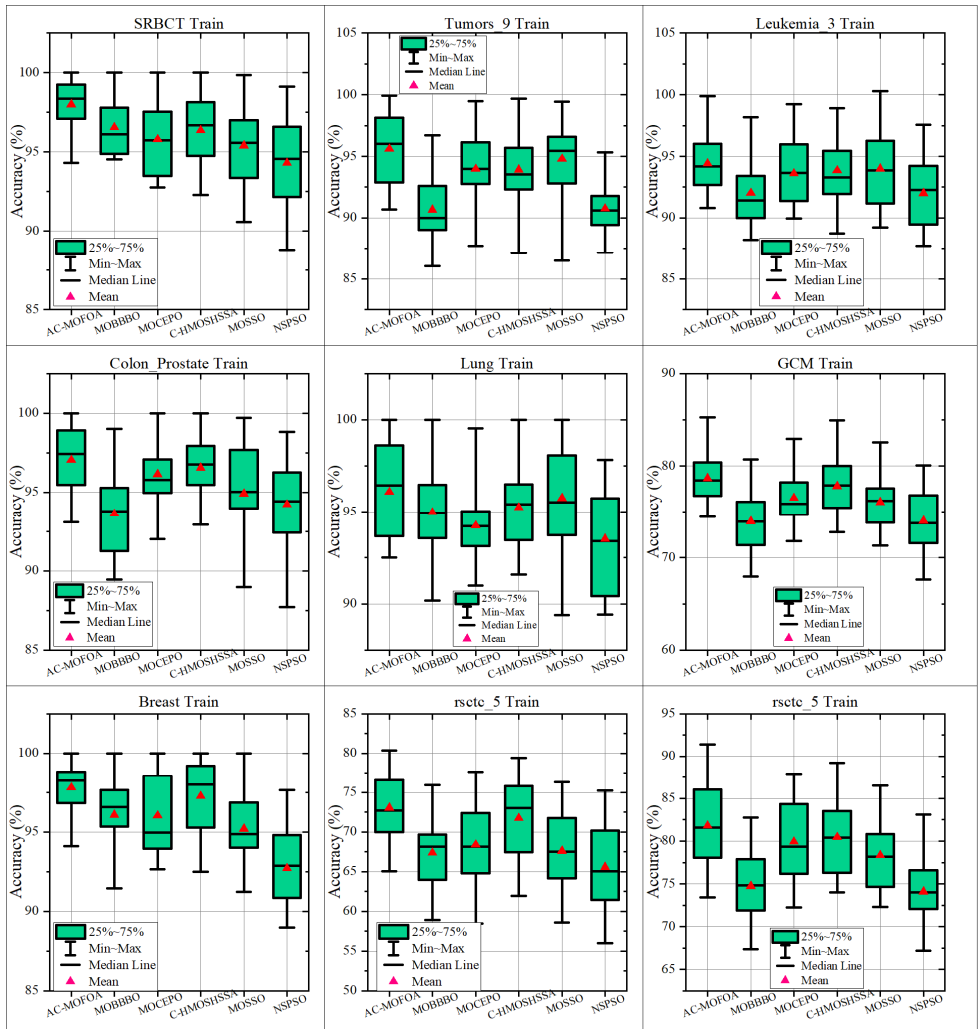

**Fig. 9** Comparing the distributions of classification accuracy of AC-MOFOA solutions on the train set

## 5.3 Performance evaluation based on Hypervolume and Student t-test

To evaluate the efficiency and effectiveness of AC-MOFOA compared to other multi-objective algorithms, Student T-test was performed on normalized values of Hypervolume. To perform the t-test, the Hypervolume values were first calculated in 50 independent executions for the MOSSO, MOCEPO, C-HMOSHSSA, MOBBBO, NSPSO, and AC-MOFOA algorithms. Next, the t-test was performed with a significance level of 0.05. Tables 6 and 7 show the t-test results in both test and train modes, respectively. Comparisons were made from left to top, respectively. Superior, similar, and worse performance of the AC-MOFOA compared to the corresponding algorithm are displayed by "+", "=", and "-", respectively.

Table 6 T-test of hypervolume ratios in the *test* data

| Datasets | MOCEPO | MOBBBO | C-HMOSHSSA | MOSSO | NSPSO |
|---|---|---|---|---|---|
| SRBCT | + | + | + | + | + |
| Tumors_9 | + | + | + | + | + |
| Leukaemia3 | + | + | + | + | + |
| Colon_Prostate | + | + | + | + | + |
| Lung | = | + | + | + | + |
| GCM | + | + | = | + | + |
| Breast | + | + | = | + | + |
| Rsctc_5 | + | + | = | + | + |
| Rsctc_6 | + | + | = | + | + |

As can be seen from Table 6, the AC-MOFOA results in all test datasets are significantly superior to the three methods of MOBBBO, NSPSO, and MOSSO. Moreover, MOCEPO provided similar results to AC-MOFOA only in the Lung database and showed worse performance in other cases. AC-MOFOA could also achieve superior results over C-HMOSHSSA in 5 out of 9 datasets and had similar results in 4 datasets.

Table 7 T-test of hypervolume ratios in the *train* data

| Datasets | MOCEPO | MOBBBO | C-HMOSHSSA | MOSSO | NSPSO |
|---|---|---|---|---|---|
| SRBCT | = | + | + | + | + |
| Tumors_9 | + | + | + | + | + |
| Leukaemia3 | = | + | + | + | + |
| Colon_Prostate | + | + | + | + | + |
| Lung | = | + | + | + | + |
| GCM | + | + | = | + | + |
| Breast | + | + | = | + | + |
| Rsctc_5 | + | + | = | + | + |
| Rsctc_6 | + | + | + | + | + |

Table 7 shows the results of the t-test on the Hypervolume values of the train data. According to the results of the t-test, it can be concluded that AC-MOFOA outperformed other methods in most datasets. AC-MOFOA showed similar results with MOCEPO and C-HMOSHSSA in only a few cases. For example, we can refer to three datasets of Lung, SRBCT, and Leukemia3 in which AC-MOFOA had a similar performance to MOCEPO.

## 5.4 Investigating the time and space complexity

Time and space complexity analysis is commonly used to analyze the performance of computer algorithms. Therefore, in this section, we discuss AC-MOFOA complexity analyses. By analyzing the AC-MOFOA algorithm, it can be seen that the two steps of 8 and 11 have the highest time complexity among all the steps of the algorithm. In step 8, a non-dominated sorting operation is performed with the time order of $O(MN^2)$, assuming that $M$ is the number of objectives and $N$ is the number of trees. Also, based on a study conducted by Jensen [82], with an optimal implementation, the time complexity of $O(N \log^{M-1} N)$ can be achieved. In Step 11, crowding-distance is calculated based on time complexity of $O(MN \log N)$. Based on the mentioned points, the time complexity of AC-MOFOA will be from the $O(MN^2)$ in the normal implementation and $O(N \log^{M-1} N)$ in the optimal implementation. Jensen [82] reported that NS-based algorithms in the normal implementation are from $O(MN^2)$.

AC-MOFOA has two memory sections that are used to store Forest and Candidate Population, respectively. Assuming that the dimensions of each tree are $k$, the space complexity required to store the Forest will be from $O(N \times k)$. The space complexity of Candidate Population will also be from $O(N_{cand.} \times k)$. Therefore, it can be concluded that the general space complexity of AC-MOFOA is from $O(\max(N, N_{cand.}) \times k)$.

Figure 10 presents the results of comparing the mean execution time of AC-MOFOA with other multi-objective algorithms. The mean execution time of algorithms is calculated based on the 50 independent executions time on each of the datasets.

Analyzing the AC-MOFOA execution time shows that this method had a shorter execution time than MOCEPO, MOBBBO, C-HMOSHSSA, and NSPSO. Due to its single-parent structure, AC-MOFOA only needs to select one parent among the Pareto Front members. Thus, it spends less time in the parent-selection step. Also, AC-MOFOA only needs to make limited changes in the structure of the parent tree to generate new trees, resulting in the reduced computational overhead of global and local seeding operators. Furthermore, implementing the chaos theory on local and global seeding operators increases the AC-MOFOA convergence speed. At the same time, it maintains the necessary diversity needed for effective search of the problem space. As a result of this, the execution time of AC-MOFOA can be reduced compare to the traditional methods. Among the methods compared, the MOSSO algorithm had a shorter execution time than AC-MOFOA in some scenarios because of using the Archive structure to maintain the Pareto front and being a single parent.

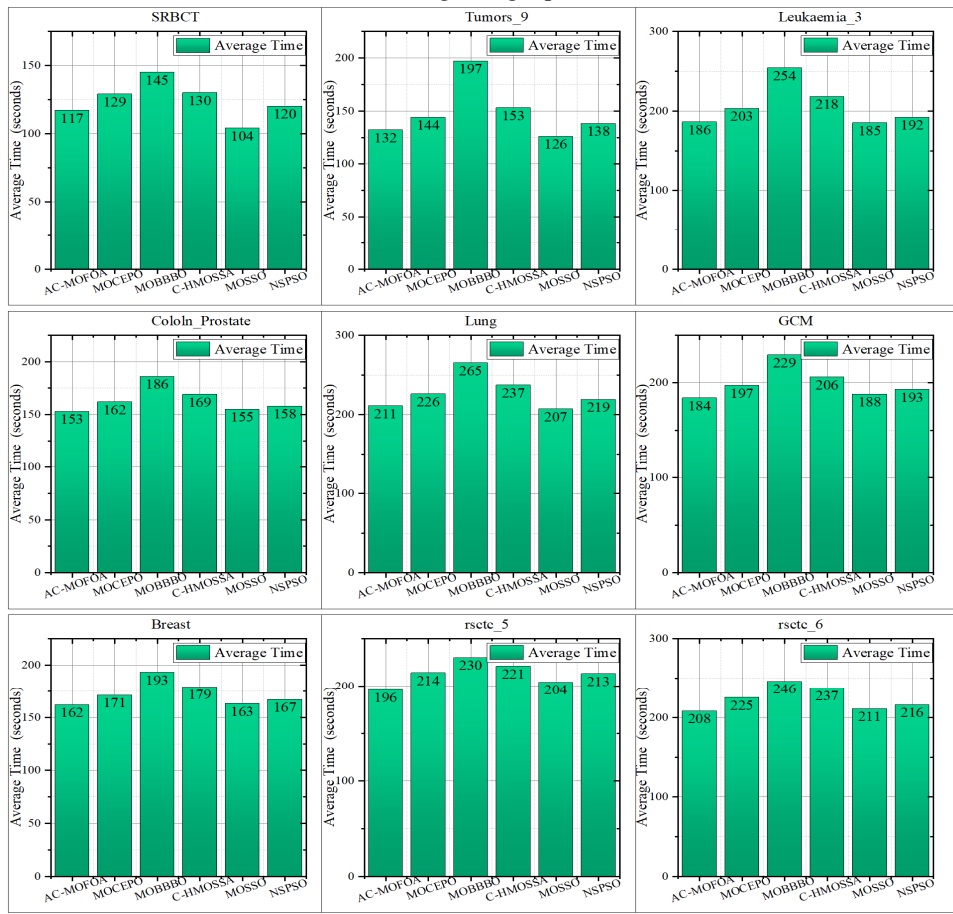

**Fig.10** Comparison of the mean execution time of AC-MOFOA with other multi-objective algorithms (in seconds)

### 5.5 Studying the results of the ensemble classifier

In this section, we analyze the results of the Ensemble classifier made from members of the Pareto Front. The comparison results of the proposed Ensemble classifier with the three ensemble learning methods (i.e., *Adaboost, Bagging, and Random Forest*) are shown in Table 8. Analysis of the results suggests that the proposed Ensemble classifier in all datasets could improve the classification accuracy. The proposed method also outperformed the other methods in 7 datasets. For example, in Breast dataset, the proposed method could improve the classification accuracy by more than 30% compared to the non-optimized KELM. In fact, due to the optimization of the RBF kernel ($\gamma$) and ELM ($C$) function parameters, AC-MOFOA could improve the classification efficiency of each of the Pareto Front KELMs simultaneously with reducing the number of genes. Accordingly, the combination of improved KELMs could provide better results compared to conventional ensemble learning methods. However, the proposed method in two datasets of Leukemia3 and Lung yielded less accurate solutions than the Random Forest and Bagging methods.

Table 8 Comparison results of proposed Ensemble Classifier

| Datasets | KELM | Proposed Ensemble Classifier | Random Forest | Adaboost | Bagging |
|---|---|---|---|---|---|
| SRBCT | 88.23 | **98.65** | 91.97 | 95.18 | 94.59 |
| Tumors_9 | 78.98 | **95.13** | 92.67 | 93.15 | 90.6 |
| Leukaemia3 | 97.22 | 97.86 | **97.93** | 96.74 | 94.13 |
| Colon_Prostate | 95.74 | **96.92** | 96.08 | 95.27 | 95.78 |
| Lung | 90.24 | 97.6 | 92.14 | 91.61 | **98.04** |
| GCM | 70.53 | **79.11** | 74.24 | 72.05 | 76.19 |
| Breast | 68.04 | **97.45** | 82.51 | 88.36 | 91.31 |
| Rsctc_5 | 69.66 | **73.52** | 70.48 | 71.66 | 71.98 |
| Rsctc_6 | 81.523 | **83.94** | 79.73 | 80.73 | 82.64 |

## 5.6 Further discussion

Based on the obtained results, the proposed method could successfully achieve the set objectives (i.e., reducing the dimensions of Microarray datasets, increasing the accuracy of the classification, and constructing the Ensemble Classifier). The considered Multi-Filter includes five filter methods, including IG, Fisher-score, mRMR, CFS, and ReliefF. The selected combination includes Univariate and Multivariate Methods. In this system, Univariate methods rank genes based on their individual relation with output. On the other hand, Multivariate methods rank genes based on their intra-relations and relation with output. By reducing the bias of single-filter methods, Multi-filter could select prominent and influential genes of the dataset and reduce dimensions of them to an acceptable level. Additionally, Multi-filter considers the dependency of each gene with output and redundancy between genes. Furthermore, AC-MOFOA simultaneously reduced the size of the gene subset and improved classification accuracy using the *NS*, crowding distance, chaos theory, adaptive local, and adaptive global seeding operators.

The use of adaptive local and global seeding operators enabled the AC-MOFOA to overcome the challenge of determining *LSC* and *GSC*, which leads to broader global and local search to find the optimal Pareto front in the early stages. Due to the implementation of the chaos theory concepts in local and global seeding operators, AC-MOFOA could achieve faster convergence compared to the traditional methods. Furthermore, the local seeding operator in AC-MOFOA uses the solutions of $F_1$ for generating new trees, which improves AC-MOFOA's search ability toward the optimal Pareto front. Besides, the use of *NS* and crowding distance enabled the algorithm to perform well in identifying the uniform and divers Pareto fronts. Finally, members of final Pareto front, including pairs of subset genes and KELM, were utilized to construct the Ensemble classifier. It is of note that the members of the Pareto front have the necessary diversity to construct an ensemble learner because of using different gene subsets and KELM classifiers with different optimized parameters. The results show that the proposed Ensemble classifier outperforms conventional Ensemble classification methods in terms of classification accuracy and generalizability.

## 6 Conclusion

In the present study, a hybrid method of Multi-filter and AC-MOFOA was presented to solve the problem of gene selection and construct Ensemble Classifiers for the microarray datasets. Based on experiments and results, it could successfully achieve the set objectives. The proposed multi-filter was developed by combining five filter methods, namely IG, Fisher-score, CFS, mRMR, and ReliefF, using the voting mechanism. The combination of several univariate and multivariate filter methods in constructing Multi-filter has reduced the bias effect compared to single filter mode and, as a result, the proposed Multi-filter as a pre-processing step could reduce the dimensions of the data and increase the classification accuracy. In the second step of the proposed hybrid method, AC-MOFOA was presented using the concepts of *NS*, crowding distance, chaos theory, and adaptive operators. The second step of the proposed method's objectives was selecting the quasi-optimal subset of genes, optimizing the KELM classification parameters, and increasing the classification accuracy. AC-MOFOA uses *NS* to identify the Pareto front and rank the solutions. In addition, using the crowding distance, the ability to direct the AC-MOFOA search to the less crowded areas of the Pareto front increased, leading to a more uniform final Pareto front. Using chaos theory and adaptive local and global seeding, AC-MOFOA could improve its global and local search without needing to set the *LSC* and *GSC* parameters. Moreover, utilizing the logistic map as a chaos function in global and local seeding operators, the convergence speed and search diversity of AC-MOFOA have been improved. Finally, in the third step of the proposed method, the Ensemble Classifier model was formed using the final Pareto front KELM classifiers. To evaluate the effectiveness and efficiency of the proposed solution, the results were compared with five hybrid multi-objective gene selection methods of MOSSO, MOCEPO, C-HMOSHSSA, NSPSO, and MOBBBO on nine microarray datasets with different dimensions. Based on the results, AC-MOFOA in most datasets could simultaneously increase classification accuracy by reducing the dimensions of the datasets. Moreover, the distribution analysis of the accuracy of solutions and the number of genes selected by AC-MOFOA based on statistical criteria confirm its effectiveness. Furthermore, the hypervolume indicator and SCC analysis show that AC-MOFOA achieves better results on identifying the optimal Pareto front. Additionally, the proposed Ensemble Classifier model provides better performance for microarray data classification by increasing the accuracy of classification compared to conventional ensemble learning methods (i.e., *Adaboost, Bagging, and Random Forest*). AC-MOFOA could successfully solve the problem of gene selection and construction of the Ensemble classifier.

However, due to developments and advances in DNA microarray that have resulted in increasing dimensions and samples of such data, conventional computers will lose their ability to solve the problem of gene selection and microarray data classification. Hence, there is an urgent need to develop methods based on big data architecture. In future studies, we will review and develop scalable methods based on multi-objective metaheuristic algorithms in the big data context.